\documentclass[12pt,preprint]{aastex}

\usepackage{lscape}
\usepackage{graphicx}

\begin{document}

\title {Interstellar detection of c-C$_3$D$_2$ \footnote{Based on observations carried out with the IRAM 30m Telescope. IRAM is supported by INSU/CNRS (France), MPG (Germany) and IGN (Spain).}}


\author{S. Spezzano\altaffilmark{1,2}\footnote{Member of the International Max
  Planck Research School (IMPRS) of Astronomy and Astrophysics at the
  Universities of Bonn and Cologne}, S. Br\"unken\altaffilmark{1},
  P.Schilke\altaffilmark{1}, P.Caselli\altaffilmark{3},
  K.~M.~Menten\altaffilmark{2}, M.~C.~McCarthy\altaffilmark{4},
  L. Bizzocchi\altaffilmark{5}, S.~P. Trevin\~{o}-Morales\altaffilmark{6}, Y. Aikawa\altaffilmark{7} and
S. Schlemmer\altaffilmark{1}}
\altaffiltext{1} {I. Physikalisches Institut, Universit\"at zu
  K\"oln, Z\"ulpicher Str. 77, 50937 K\"oln, Germany} \altaffiltext{2}
{Max-Planck-Institut f\"ur Radioastronomie, Auf dem H\"ugel 69, 53121
  Bonn, Germany} \altaffiltext{3}{School of Physics and Astronomy, University of Leeds, Leeds LS2 9JT, UK}\altaffiltext{4}{Harvard-Smithsonian
Center for Astrophysics, 60 Garden St., Cambridge, MA 02138, and
School of Engineering \& Applied Sciences, Harvard University, 29
Oxford St., Cambridge, MA 02138} \altaffiltext{5}{Centro de Astronomia e Astrof\'isica, Observat\'orio Astron\'omico de Lisboa, Tapada da Ajuda, 1349-018 Lisboa, Portugal} \altaffiltext{6}{IRAM, 18012, Granada, Spain}\altaffiltext{7}{Department of Earth and Planetary Sciences, Kobe University, Kobe 657-8501, Japan}


\begin{abstract}
We report the first interstellar detection of c-C$_3$D$_2$. The doubly
deuterated cyclopropenylidene, a carbene, has been
detected  toward the starless cores TMC-1C and L1544 using the IRAM
30m telescope. The $J_{K_a,K_c} = 3_{0,3} - 2_{1,2}$, $3_{1,3} - 2_{0,2}$, and
$2_{2,1} - 1_{1,0}$ transitions of this species have been observed at 3 mm in both sources.
The expected 1:2 intensity ratio has been found in
the 3$_{0,3}$ - 2$_{1,2}$ and  3$_{1,3}$ - 2$_{0,2}$ lines, belonging
to the para and ortho species respectively. We also observed lines of the main species,
c-C$_3$H$_2$, the singly deuterated c-C$_3$HD, and the species with
one $^{13}$C off of the principal axis of the molecule, c-H$^{13}$CC$_2$H. The lines of c-C$_3$D$_2$ have been observed
with high signal to noise ratio, better than 7.5$\sigma$ in TMC-1C
and 9$\sigma$ in L1544. The abundance of doubly deuterated cyclopropenylidene
with respect to
the normal species is found to be  (0.4 - 0.8)\% in TMC-1C and (1.2 -
2.1)\% in L1544. The deuteration of this small hydrocarbon ring is analysed
with a comprehensive gas-grain model, the first including doubly deuterated species. The observed abundances of c-C$_3$D$_2$ can be explained solely by gas-phase processes, supporting the idea that c-C$_3$H$_2$ is a good indicator of gas-phase deuteration.
\end{abstract}





\section{Introduction}

Investigating deuterium chemistry
is useful to put constraints on the ionization fraction, temperature,
density and thermal history of dense molecular
clouds \citep{Guelin_1977, Caselli_2002, Cazaux_2011, Taquet_2012}. 
The observations of multiply deuterated molecules in space, e.g. \citep{Ceccarelli_2007} and 
references therein, have shown
the necessity to reexamine some reaction rates in chemical networks
\citep{Roberts_2002}, elemental D/H ratio in cold dense gas \cite{Roueff_2007} and the density structure in sources such as L1544 and $\rho$~Oph D
\citep{Roberts_2004}, as well as the effects of accretion on grains
\citep{Roberts_2000}, possible effects of internal dynamical motion
\citep{Aikawa_2005}, and the evolution of ice mantles in dense clouds and cores 
\citep{Cazaux_2011, Taquet_2012}.

The first multiply deuterated interstellar molecule detected has been
D$_2$CO almost twenty years ago \citep{Turner_1990}. Since then, the study
of deuterated molecules in the ISM has rapidly increased as they have been proven to be
a unique observational probe of the early stages in low-mass star
formation. Multiply deuterated species such as triply deuterated
ammonia have been detected with a surprisingly high
abundance ratio of 10$^{-4}$ with respect to their fully hydrogenated
forms \citep{Lis_2002}. By comparing this ratio with the elemental
D/H ratio (1.65$\times10^{-5}$, Linsky et al. 1993) it is easily seen that there is a
remarkable enrichment in deuterium in this and
other molecules.
 In IRAS 16293-2422 Ceccarelli et al. (1998)
measured D$_2$CO/H$_2$CO = 5\%. In the same source Parise et al. (2004)
measured CD$_3$OH/CH$_3$OH = 1.4\%. 
This enrichment in deuterium has mainly been explained by the exothermicity of the H-D exchange reactions, and by the depletion of
CO and O onto
dust grains \citep{Dalgarno_1984, Caselli_1999}, which enhances
the abundance of the multiply deuterated forms of H$_3^+$ and, upon
dissociative recombination, the D/H ratio.
Roberts et al. (2003) calculated D/H abundance ratios close to 0.3
(20,000 times the cosmic value), in regions with large amount of CO freeze-out.

Despite extensive studies of nearly 30 deuterated molecules in
the interstellar gas, the processes regulating their formation are not completely understood, in 
particular the relative importance of gas-phase versus grain-surface
chemistry. For example, simple gas phase chemical models do not
reproduce the deuterium enrichment observed in some of these
species. In particular, the formation of methanol on grain mantles is presented as an
explanation for its enhanced deuteration \cite{Parise_2006}; for H$_2$CO instead, the gas-phase reaction path cannot
be excluded. In fact, in the Orion Bar photodissociation region HDCO seems to be
formed solely in the gas phase \cite{Parise_2009}. Deuteration of ammonia can be reproduced with gas-phase models (e.g. Roueff et al. 2005), although NH$_3$ and its deuterated forms are also expected to form on the surface \cite{Tielens_1983}. 

The first detection of c-C$_3$H$_2$ in the laboratory \citep{Thaddeus_1985}
allowed the identification of several U-lines previously detected in
space \cite{Thaddeus_1981}, namely the strong
lines at 85338 and 18343 MHz. Since then, cyclopropenylidene has been proven to be one of the most abundant
and widespread molecules in our Galaxy. It has been observed in the diffuse gas,
cold dark clouds, giant molecular clouds, photodissociation regions,
circumstellar envelopes, and planetary nebulae \cite{Thaddeus_1985, Vrtilek_1987,
Cox_1987, Madden_1989, Lucas_2000}. Given the high abundance
of the normal species, both c-C$_3$HD
and the singly substituted $^{13}$C species (off axis) have been
observed with a good signal to noise in cold dark clouds.  Furthermore
cyclopropenylidene shows an enhancement in deuterium fractionation in
cold dark clouds, for example Gerin
et al. (1987) measured a 1:5 ratio of the 2$_{1,2}$-1$_{0,1}$ lines
of c-C$_3$HD and c-C$_3$H$_2$ in TMC1.
The reactions which lead to such a high deuteration are still poorly
understood. Some rates of reactions which may be involved in the
formation of deuterated c-C$_3$H$_2$ have been measured by Savi\'c et
al. (2005) but to our knowledge they have not been included in models
so far.
Last year the centimeter and millimeter wavelength spectra of doubly
deuterated c-C$_3$H$_2$ have been measured in the laboratory \citep{Spezzano_2012}, allowing for the
first time a search for c-C$_3$D$_2$ in space. Like its fully
hydrogenated counterpart, c-C$_3$D$_2$ presents
the spectrum of an oblate asymmetric top with b-type
transitions. Furthermore c-C$_3$D$_2$ shows a deuterium quadrupole
splitting resolvable at very low $J$. Given the presence of
two equivalent off axis bosons, it has {\it ortho} and {\it para} symmetry species
with relative statistical weight of 2:1. 

Unlike several of the six known multiply deuterated species observed
in the radio band (D$_2$H$^+$, CHD$_2$OH, NHD$_2$, D$_2$CO, D$_2$S,
and D$_2$CS), c-C$_3$H$_2$ is believed to form solely by gas phase reactions \cite{Park_2006}.
The interplay between the gas phase and grain surface reactions in
the deuteration of interstellar molecules is not clear so far, partially
because there are not many probes available for testing the models:
c-C$_3$H$_2$ is an ideal molecule for this purpose because of its
easily observable transitions and because it has
the possibility of double deuteration. Assuming c-C$_3$D$_2$ is formed in the gas-phase like its fully
hydrogenated counterpart, cyclopropenylidene will be a unique probe
for the deuteration processes happening in the gas-phase.
Furthermore, since c-C$_3$H$_2$ is, in terms of cloud evolution, an ”early-type” molecule
\cite{Herbst_1989}, it is a particular useful tool to investigate early stages of a molecular cloud. 
This makes observations of its deuterated forms particularly important to test time-dependent chemical codes which include deuteration processes.

Here we report on the positive detection of three emission lines of
c-C$_3$D$_2$ in the 3 mm band, namely the $J_{K_a,K_c} = 3_{0,3} - 2_{1,2}$, $3_{1,3} -
2_{0,2}$, and
$2_{2,1} - 1_{1,0}$ transitions, towards TMC-1C and
L1544: to our knowledge this is the first search for doubly deuterated
cyclopropenylidene undertaken. 
\footnote{A recent abstract from S. Takano et al. at the Workshop on
Interstellar Matter 2012 in Sapporo, Japan, mentions the detection of
c-C$_3$D$_2$ towards L1527 in the framework of the Nobeyama 45m
telescope survey at 3 mm.}

\section{Observations}


The observations have been carried out from 2012 September 28
until October 2 at
the IRAM 30m telescope, located in Pico Veleta (Spain), towards the starless
cores TMC-1C and L1544.
The choice of the sources has been made on the basis of two simple
criteria: the abundance of the normal species (c-C$_3$H$_2$) and a high deuterium
fractionation.
Both sources are in the Taurus Molecular Cloud, one of the closest
dark cloud systems and low-mass star
forming regions in our Galaxy. L1544 is a perfect test bed to investigate the
initial conditions of protostellar collapse: its structure is consistent with a contracting Bonnor-Ebert
sphere, with central densities of about 10$^7$ cm$^{-3}$ and a peak
infall velocity of $\simeq$0.1 km s$^{-1}$ at about 1000 AU from the
center (e.g. Keto \& Caselli 2010). Its centrally concentrated
structure and measured kinematics suggests that this is a pre-stellar
core at a late stage of evolution, toward star formation. TMC-1C is a
relatively young core, with evidence of accreting material towards a
core and immersed in a cloud with densities higher than those
surrounding the L1544 core \cite{Schnee_2007}.

The coordinates that were used are $\alpha _{2000}$ = 04$^h$41$^m$16$^s$.1
 $\delta _{2000}$ = +25$^\circ$49$'$43$''$.8 for TMC-1C, and  $\alpha _{2000}$ =
05$^h$04$^m$17$^s$.21  $\delta _{2000}$ = 25$^\circ$10$'$42$''$.8 for L1544. In the case
of TMC-1C these are the same coordinates as reported by Bell et al. (1988)
and Gerin et al. (1987), while in the case of L1544 they correspond to the coordinates of
the peak of the 1.3 mm continuum dust emission from Ward-Thompson et al. (1999). In both cores we observed two lines of
c-C$_3$H$_2$, one line of c-H$^{13}$CC$_2$H (off axis), two lines of c-C$_3$HD and three lines
of c-C$_3$D$_2$, using three different tuning settings.
A summary of the observed lines is reported in Table 1. The EMIR receivers in the E090 configuration were employed, and observations were performed in a frequency switching mode with a
throw of $\pm4.3$ MHz. 
All four EMIR sub-bands were connected to the FTS spectrometer set to
high resolution mode; this delivered a final spectrum with 50 kHz
channel spacing (corresponding to 0.15 km s$^{-1}$ at 3 mm) and a total of 7.2 GHz of spectral coverage (nominal bandpass of 1.8 GHz per sub-band).
Telescope pointing was checked every two hours on Jupiter and was found accurate to ~3-4 arcsec.




\section {Results}
Lines of the isotopologues of  c-C$_3$H$_2$ listed in Table 1 have
been detected in both sources with very high signal to noise
ratio. A selection of spectra of c-C$_3$H$_2$ and isotopologues in
TMC-1C and L1544 is shown in Figure 1.
Table 1 lists the observed line parameters.
Even the weakest line of the doubly deuterated species, 2$_{2
  1}$ - 1$_{1 0}$ at 108 GHz, is
detected at a 7.5$\sigma$ level in TMC-1C (T$_{mb,rms}$ = 2.5 mK), and
at a 9$\sigma$ level in L1544 (T$_{mb,rms}$ = 4.6 mK).
The GILDAS\footnote{http://www.iram.fr/IRAMFR/GILDAS} software \cite{Pety_2005} was employed for the data
processing: high order polynomials had to be used for baseline
subtraction given the strong baseline produced by the frequency
switching observing mode. The column densities and optical depths given in Table 1 were
  calculated using the expressions given in the Appendix. As was already pointed out by Bell et al. (1988), c-C$_3$H$_2$
shows two velocity components towards TMC-1C, one more intense at 6 km
s$^{-1}$ and one less intense at 5.4 km s$^{-1}$, see Figure 1. There is a hint of
detection of the component at 5.4 km s$^{-1}$ also in c-H$^{13}$CC$_2$H, but no clear presence in the deuterated
species. Assuming that all lines have the same excitation
temperature in both components, we expect for the component at 5.4 km s$^{-1}$
a line intensity of 0.06 K for c-C$_3$HD (3$_{0 3}$ - 2$_{1 2}$) and
0.013 K for c-C$_3$D$_2$ (3$_{1 3}$ - 2$_{0 2}$). Comparing these
estimates with the noise level in our spectra (0.007 K for c-C$_3$HD
and 0.002 K for  c-C$_3$D$_2$) we can say that the lower velocity component is absent
in the deuterated species of cyclopropenylidene: this behavior may suggest that the
lower velocity component traces a hotter region, where the deuterated
molecules are not present in detectable amounts.

\subsection{Excitation}
The observed line intensity ratios of c-C$_3$D$_2$ pose the question whether local thermal equilibrium is a valid assumption for its excitation. Only for \emph{para}-C$_3$D$_2$ more than one
optically thin line was detected.  The ratios of the integrated
3$_{0,3}$-2$_{1,2}$/2$_{2,1}$-1$_{1,0}$ lines should be 4.3 and 3.3
for TMC-1C (T$_{\rm ex}$ = 7 K) and L1544 (T$_{\rm ex}$ = 5 K),
respectively, assuming thermalization of both lines to their assumed
excitation temperature, while they are 1.5 and 1.35. To gain insight into the excitation of the lines we
performed radiative transfer calculations with RADEX
\cite{Vandertak_2007}. Collision rates of C$_3$H$_2$ with H$_2$ calculated by
Chandra et al. (2000) and supplied by the LAMDA database
\cite{Schoier_2005} were used together with molecular constants of
\emph{para}-C$_3$D$_2$. Calculations were done on a grid in density of
molecular hydrogen, n(H$_2$), and column density of \emph{para}-C$_3$D$_2$, N(\emph{para}-C$_3$D$_2$), and the results are shown in Figure
2. It is in principle possible to read the column density of c-C$_3$D$_2$ from Figure 2, by
knowing the density of molecular hydrogen. The nominal observed line ratios would, particulary for L1544, result
in very high column densities.  The lines would be very
subthermally excited and very optically thick.  However, given the
fact that the observed line strengths run parallel to each other, a
small change in the observed value would result in a substantial
change in the ratio. The line
intensities and $\chi ^2$ contours of both lines show that the line strengths are very
close to each other along a diagonal line spanning a large range of
densities and column densities (filling factors smaller than unity
would move this line to the right in the plot), so that much smaller
values of column densities and opacities are also in agreement with
the data.  The predicted excitation temperatures  tend to be rather
low, in the range 3--3.5~K, which introduces a considerable
uncertainty in the column density determinations. Since the
transitions used for the excitation calculation for c-C$_3$HD and c-C$_3$D$_2$
have similar upper state energy, the excitation behavior will naturally
be degenerated. Therefore, in the
absence of data for more transitions for c-C$_3$D$_2$,  c-C$_3$HD or
c-H$^{13}$CC$_2$H, it is difficult to derive conclusive results. Despite these uncertainties, we have assumed local thermodynamic
equilibrium with $T_{ex}$ values as described in the Appendix.

\section{Discussion}
In addition to the observation of two lines of  c-C$_3$H$_2$, one line
of c-H$^{13}$CC$_2$H (with the $^{13}$C off of the molecular axis), and two lines of c-C$_3$HD, we also claim the detection of three lines of c-C$_3$D$_2$ in both TMC-1C and L1544.
This first interstellar detection of  c-C$_3$D$_2$  is validated by the following reasons:
\begin{itemize}
\item{We detected all favorable transitions of c-C$_3$D$_2$ available in the
  covered frequency range.} 
\item{The rest frequencies employed have laboratory accuracy
  \cite{Spezzano_2012}, and in both sources the line shapes and velocities are
  in agreement with each other and with those observed for more
  abundant isotopologues (see Figure 1).}
\item{The intensities of the  c-C$_3$D$_2$ lines are consistent with
  what is expected from the deuteration of the ring in these sources,
  i.e.  c-C$_3$H$_2$/c-C$_3$HD is consistent with
  c-C$_3$HD/c-C$_3$D$_2$, as will be discussed below.}
\end{itemize}

In Table 2 we present the relative abundances of the deuterated
species with respect to the hydrogenated ones for cyclopropenylidene over
the 27$''$ beam for TMC-1C
and L1544 obtained from this work, and also for H$_2$CO,
HCO$^+$, N$_2$H$^+$ and NH$_3$ obtained from previous work. The abundance of doubly deuterated cyclopropenylidene
with respect to
the normal species is (0.4 - 0.8)\% in TMC-1C and (1.2 - 2.1)\% in
L1544. This interval has been determined considering the
  differences in $N_{tot}$ obtained from different lines. The
deuteration of c-C$_3$H$_2$ follows the same trend observed for other molecules in both sources. It is interesting to note that the ratios [c-C$_3$D$_2$]/[c-C$_3$HD] and
[c-C$_3$HD]/[c-C$_3$H$_2$] are quite similar in both sources.
 We calculated the D/H ratio of c-C$_3$H$_2$ in prestellar cores using the
 network model of Aikawa et al. (2012). For the physical structure of the core, we adopt
 the collapsing core model of Aikawa et al. (2005; the $\alpha$=1.1
 model) and also a static model of L1544 from Keto \& Caselli
 (2010). For CO depletion factors consistent with those of the two
 objects, i.e. $f_D$ = 3.8 for TMC-1C \cite{Schnee_2007} and $f_D$ =
 14 for L1544 \cite{Crapsi_2005}, the calculated column density ratio of
 c-C$_3$D$_2$/c-C$_3$H$_2$ is $\sim10^{-2}$, consistent with the
 observed value of 0.6\% for TMC-1C and 1.5\% for L1544. There is no need for any deuterium
 fractionation reactions of c-C$_3$H$_2$ on grain surfaces to account
 for the observed D/H ratio: the deuteration of cyclopropenylidene can
 be explained solely by gas-phase reactions.
 The main route of formation of deuterated
 cyclopropenylidene is the successive deuteration of the main species
 via reaction with H$_2$D$^+$, D$_2$H$^+$, and D$_3^+$. An example of
 the reaction scheme is sketched in Figure 3, considering only
 H$_2$D$^+$ as reaction partner.
The depicted cycle of reactions starts with
c-C$_3$H$_2$ and H$_2$D$^+$, producing in the first step c-C$_3$HD and
subsequently c-C$_3$D$_2$. The same reactions happen with D$_2$H$^+$
and D$_3^+$. The overall process is a series of two
reactions: the proton-deuteron transfer (slow step, red arrows), and
the subsequent dissociative recombination with electrons (fast step,
blue arrows). The presence of this deuteration cycle results in a time
dependent deuterium fractionation. Assuming low levels of deuteration
at the start, it is expected that this level increases as a function
of time, reaching a stationary level after some time. Other deuteration processes, e.g. the formation of c-C$_3$HD from the reaction of C$_3$H$^+$ with HD, were found to be negligible. The D/H ratio of cyclopropenylidene is, therefore, directly related to that of H$_3^+$, the main deuterium donor in dark interstellar clouds.
Recently Huang \& Lee (2011) have calculated highly accurate
spectroscopic constants for $^{13}$C and D isotopologues of
c-C$_3$H$_3^+$ in order to guide the laboratory and astronomical
search. Since these species are intermediates in the formation of isotopic
species of c-C$_3$H$_2$, their detection would be useful to put more
constraints on the models.

Doubly deuterated cyclopropenylidene appears to be a very interesting
probe for the earliest stages of star formation. Its formation
mechanism puts important constraints on gas-phase deuteration models,
and suggests the possibility of using  c-C$_3$D$_2$ as a chemical
clock. Furthermore the brightness of the $J_{K_a,K_c} = 3_{1,3} -
2_{0,2}$ line of c-C$_3$D$_2$ at 97 GHz in L1544 would allow
to make an on the fly map (OTF) of the core in a reasonable amount of
time, for the first time for a doubly deuterated molecule. During our
observations, the line was
observed with a S/N of more than 6 after just 30 minutes. 
Emission from C$^{34}$S is spread over 10
arcmin$^2$ in L1544 \cite{Tafalla_1998}.
We estimated that an OTF map of 6
arcmin$^2$ with a noise level of 20 mK at the IRAM 30m telescope with the
EMIR receivers would be possible in 14 hours, allowing to
map at the same time c-C$_3$H$_2$ and c-C$_3$D$_2$. 
By mapping the core it will be possible to locate the deuteration
peak, and put more constraints on current gas-grains models.

\noindent {\it Acknowledgement}\\
The authors thank the referee for the useful comments.
This work has been supported by SFB956. S. Spezzano has been supported
in her research with a stipend from the International Max-Planck Research School (IMPRS) for Astronomy and Astrophysics at
the Universities of Bonn and Cologne. S. Schlemmer and S.B. acknowledge support by the Deutsche
Forschungsgemeinschaft (DFG) through project BR 4287/1-1. L.B. acknowledges support from the Science and Technology Foundation (FCT, Portugal) through the Fellowship SFRH/BPD/62966/2009, and he is grateful to the SFB956 for the travel allowance.

\section {Appendix}

The column densities and optical depths given in Table 1 were calculated using the
following expressions. The line center opacity $\tau_0$ is

\begin{eqnarray*}
\tau_0 &=& \ln\left(\frac{J(T_{ex}) -  J(T_{bg})}{J(T_{ex}) -
    J(T_{bg}) - T_{mb}}\right)\\
\end{eqnarray*}
where $J(T) = {\frac{h\nu}{k}}(e^{\frac{h\nu}{kT}}-1)^{-1}$ is the
source function in Kelvin.
The upper state column density in case of optically thin emission, and
the total column density are defined as:

\begin{eqnarray*}
N^{thin}_{up} &=&\frac{8\pi\nu^3\sqrt{\pi}\Delta\nu\tau_0}{c^3A_{ul}2\sqrt{\ln2}(e^{\frac{h\nu}{kT}}-1)}\\
N_{tot} &=& \frac{N_uQ_{rot}(T_{ex})}{g_ue^{(\frac{E_u}{kT_{ex}})}}
\end{eqnarray*}

\noindent
where k is the Boltzmann constant, $\nu$
is the frequency of the line, $h$ is the Planck constant, $c$ is the speed
of light, $A_{ul}$  is the Einstein
coefficient of the transition, $\Delta\nu$ is the full width at half
maximum, $g_u$ is the degeneracy of the upper state, $E_u$ is the
energy of the upper state, $Q_{rot}$ is the partition function of the molecule at the given
temperature $T_{ex}$. $T_{ex}$, $T_{bg}$, $T_{mb}$ are the excitation, the
background (2.7 K) and the main beam temperatures respectively, in K. 
To calculate $N_{tot}$ and $\tau$ we assumed a $T_{ex}$ of 7 K for TMC-1C and 5 K
for L1544 for all deuterated isotopologues, following Gerin et al
(1987), and 8 K for TMC-1C and 6 K
for L1544 for the main species and the $^{13}$C isotopologues as they
trace also warmer regions of the cloud. The effect of the
  excitation temperature on the derived column densities ratios in
  Table 2 was found to be small, with a change of few percent upon a variation of $\pm$1 K. By using
these expressions we assumed that the source fills the beam, and optically thin emission obeying LTE.
Since lines of c-C$_3$H$_2$ are optically thick, we derived its
total column density from the total column density of c-H$^{13}$CC$_2$H assuming  a
$^{12}$C/$^{13}$C ratio of 77, determined by Wilson \& Rood (1994)
from H$_2$CO and CO as a function of distance from the Galactic
Center, $N_u$ and $\tau$ were calculated backwards.

\begin{landscape}
\tabletypesize{\scriptsize}
\begin{deluxetable}{ccccccccccccccc}
\singlespace
\tablewidth{0pt}
\tablecaption{Observed line parameters} 
\startdata
\tableline \tableline
Molecule & Transition & Frequency & Ref.$^d$& E$_{up}$ & T$_{mb}$&rms&
W&B$_{eff}$&$\theta_{MB}$ &V$_{LSR}$&$\delta V$ & N$_u^a$ & N$_{tot}$$^{b,c}$& $\tau ^c$ \\
&(ortho/para)&(GHz)&& (cm$^{-1}$)&(K)&mK& (K km s$^{-1}$)&\%&(arcsec)&(km s$^{-1}$)& (km s$^{-1}$)& ($\times 10^{11}$ cm$^{-2}$)& ($\times 10^{12}$ cm$^{-2}$) &\\
 \hline
 \bfseries {TMC- 1C}\\
c-C$_3$H$_2$ &  2$_{1 2}$ - 1$_{0 1}$ (o) & 85.338 &1& 4.48 & 2.91&7&1.05(1)&81&29& 5.996(2) &0.338(4) &25(1) &22(1) &1.887\\
&  2$_{1 2}$ - 1$_{0 1}$ (o)& 85.338 &1& 4.48 & 1.27 &7&0.414(9)&81&29&5.361(4) & 0.307(8) & && \\
&  3$_{2 2}$ - 3$_{1 3}$ (p) & 84.727 &1& 11.21 & 0.16&7&0.047(2)&81&29& 5.984(6) &0.27(2) &3.5(2) &22(1)  &0.148 \\
c- H$^{13}$CC$_2$H & 2$_{1 2}$ - 1$_{0 1}$ & 84.185 &2& 4.40
&0.22 &7&0.060(3)&81&29& 5.977(7) &0.26(1) & 0.52(2) & 0.62(3) & 0.056\\
c-C$_3$HD & 2$_{1 1}$  - 1$_{1 0}$ & 95.994 &2&5.25 &0.1&7&
0.033(2)&80&27& 6.03(1) & 0.33(3) &1.8(1) & 2.8(2) & 0.026\\
 & 3$_{0 3}$ - 2$_{1 2}$ & 104.187 &2& 7.54 & 0.34&7&0.091(2) &79&25&
 6.034(3) &0.250(6) & 0.63(1) &1.1(3) & 0.092\\
c-C$_3$D$_2$ & 3$_{0 3}$ - 2$_{1 2}$ (p) & 94.371 &3& 6.84 &
0.04 &2&0.009(1)&80&27& 6.07(1)& 0.23(2)&0.063(5) &0.17(1)&0.010 \\
 & 3$_{1 3}$ - 2$_{0 2}$ (o) & 97.761 &3& 6.87 & 0.07 &2&0.017(1)&80&26&
 6.062(5)& 0.233(9) & 0.11(4)& 0.15(6)&0.018\\
 & 2$_{2 1}$ - 1$_{1 0}$ (p) & 108.654 &3& 5.49 &0.02&2&0.006(1)&78&24&6.06(1) & 0.25(3) &0.032(4) &0.09(1)& 0.005\\
\hline
\bfseries {L1544} \\
c-C$_3$H$_2$ &  2$_{1 2}$ - 1$_{0 1}$ (o) & 85.338 &1& 4.48
&2.44&10&1.35(1) &81&29&7.180(2) & 0.520(4) & 50(2) &37(1)& 3.579 \\
 &  3$_{2 2}$ - 3$_{1 3}$ (p) & 84.727 &1& 11.21 &
 0.21&10&0.10(1)&81&29& 7.210(8) & 0.46(1) & 4.7(2) & 37(1) & 0.172\\
c- H$^{13}$CC$_2$H & 2$_{1 2}$ - 1$_{0 1}$ & 84.185 &2& 4.40 & 0.19 &10&0.093(3)&81&29& 7.154(8)&0.44(2)  & 0.92(4) &0.96(4)& 0.096 \\
c-C$_3$HD&2$_{1 1}$  - 1$_{1 0}$ &95.994&2& 5.25 & 0.13 &10&0.065(3)&80&27& 7.17(1)&  0.48(3) & 4.1(2) &6.2(3)& 0.066 \\
& 3$_{0 3}$ - 2$_{1 2}$ & 104.187 &2& 7.54 & 0.48&10&0.238(4)
&79&25&7.181(4) & 0.468(9) & 2.1(4) &4.5(9)& 0.278\\
c-C$_3$D$_2$& 3$_{0 3}$ - 2$_{1 2}$ (p) & 94.371 &3& 6.84 & 0.07&5&0.032(2)&80&27&7.20(1) & 0.45(3) & 0.26(2) &0.77(5)& 0.035\\
 & 3$_{1 3}$ - 2$_{0 2}$ (o) & 97.761 &3& 6.87 & 0.13&5&0.059(2)&80&26&7.181(7)& 0.43(2) & 0.44(2)&0.66(2)& 0.067\\
& 2$_{2 1}$ - 1$_{1 0}$ (p)& 108.654 &3& 5.49 & 0.04 &5&0.023(2)&78&24&7.17(2) & 0.54(5) & 0.16(1) &0.45(4)& 0.020 \\
\tableline
\enddata
\tablenotetext{a}{All N$_u$ have been calculated with the optical thin
  assumption, except for c-C$_3$H$_2$, $^b${L}ocal thermodynamic
  equilibrium is assumed, $^c$T$_{ex}$ assumed is 5 K in L1544, and 7 K in TMC-1C
  for the deuterated species, and 6 K in L1544 and 8 K in TMC-1C for
  the $^{13}$C and the main species, $^d$1:Thaddeus et al. 1985. 2: Bogey et al. 1987. 3:
  Spezzano et al. 2012}
\end{deluxetable}

\end{landscape}

\begin{deluxetable}{c|cc}
\tabletypesize{\scriptsize}
\singlespace
\tablewidth{0pt}
\tablecaption{Abundance ratios of deuterated molecules in TMC-1C and L1544}
\startdata
\tableline \tableline
& \bfseries {TMC-1C}&\bfseries{L1544} \\
\hline
\textrm{[c-C$_3$D$_2$]$/$[c-C$_3$H$_2$]}& (0.4 - 0.8)\%&(1.2 - 2.1)\%\\
\textrm{[c-C$_3$D$_2$]}/\textrm{[c-C$_3$HD]}&(3 - 15){\%}&(7- 17){\%}\\
\textrm{[c-C$_3$HD]}$/$\textrm{[c-C$_3$H$_2$]}&(5 - 13){\%}&(12 - 17){\%}\\
\textrm{[D$_2$CO]}$/$\textrm{[H$_2$CO]}& - & 4\%$^a$\\
\textrm{[DCO$^+$]}$/$\textrm{[HCO$^+$]} & 2\%$^b$&4\%$^c$\\
\textrm{[N$_2$D$^+$]}$/$\textrm{[N$_2$H$^+$]} & 8\%$^b$ & 20\%$^c$\\
\textrm{[NH$_2$D]}$/$\textrm{[NH$_3$]}& 1\%$^b$ & 13\%$^d$\\
\tableline
\enddata
\tablenotetext{a}{\citet{Bacmann_2003}, $^b$\citet{Tine_2000},
  $^c$\citet{Caselli_2002}, $^d$\citet{Shah_2001}}
\end{deluxetable}

\begin{landscape}
\begin{figure}[htbp]
\centering
\includegraphics [height=80cm, width=20cm, keepaspectratio] {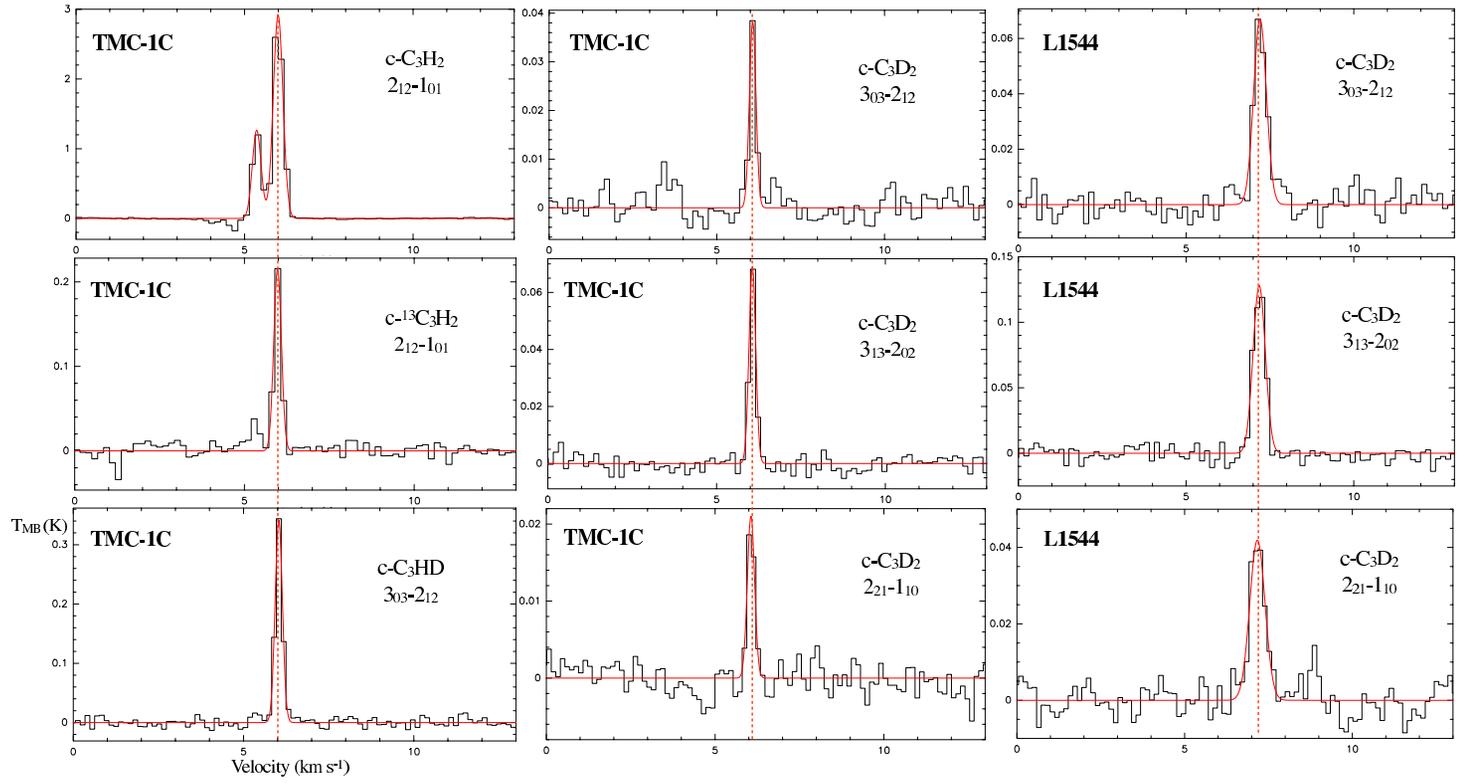}
\label{notimportant1}
\caption{Spectra of isotopic species of c-C$_3$H$_2$ observed towards TMC-1C and L1544}
\end{figure}
\end{landscape}

\begin{figure}
\centering
\includegraphics[angle=270,bb=79 300 512 700] {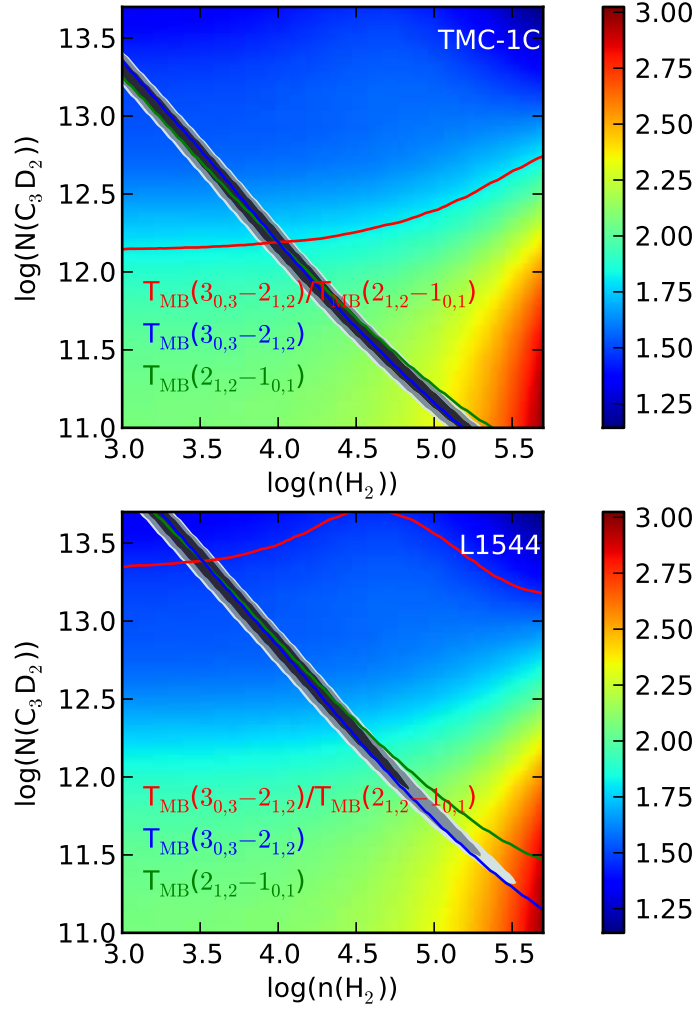}
\caption{RADEX radiative transfer calculations on the excitation of
  \emph{para}-C$_3$D$_2$. The color scale gives the 3$_{0,3}$-2$_{1,2}$/2$_{2,1}$-1$_{1,0}$ line ratio, while the observed value is drawn in as a red contour.  The observed integrated line intensities of 3$_{0,3}$-2$_{1,2}$ and 2$_{2,1}$-1$_{1,0}$ are shown as blue and green contours, respectively, and the calculated $\chi^2$ values as grey scale.}

\end{figure}

\begin{figure}[htbp]
\centering
\includegraphics [height= 20cm, width=18cm, keepaspectratio] {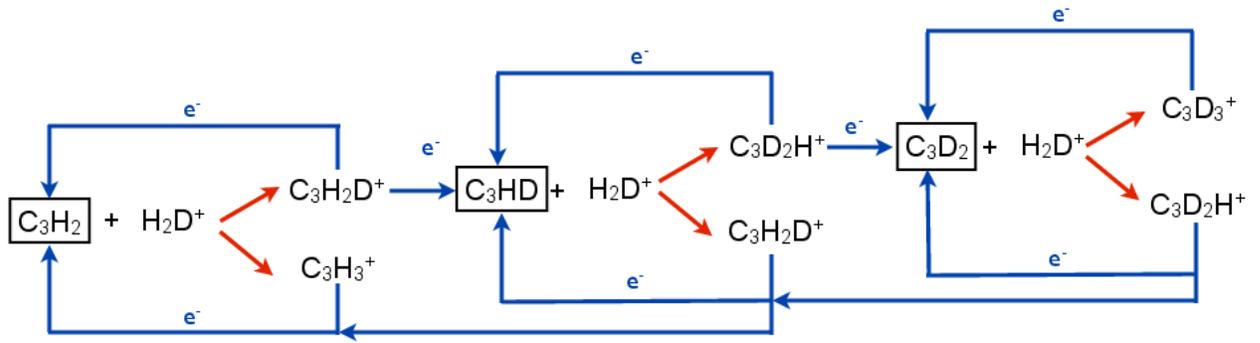}
\label{notimportant2}
\caption{Mechanism of formation of c-C$_3$D$_2$: a cycle of
  proton/deuteron transfer and dissociative recombination with electrons}
\end{figure}


\begin{thebibliography}{}

\bibitem[{{Aikawa et al.} (2005)}]{Aikawa_2005} Aikawa, Y., Herbst, E., 
Roberts, H., \& Caselli, P.\ 2005, \apj, 620, 330 

\bibitem[{{Aikawa et al.} (2012)}]{Aikawa_2012} Aikawa, Y., Wakelam, V., 
Hersant, F., Garrod, R.~T., \& Herbst, E.\ 2012, \apj, 760, 40 


\bibitem[{{Bacmann et al.} (2003)}]{Bacmann_2003} Bacmann, A., Lefloch, 
B., Ceccarelli, C., et al.\ 2003, \apjl, 585, L55 

\bibitem[{{Bell et al.} (1988)}]{Bell_1988} Bell, M.~B., Avery, L.~W., 
Matthews, H.~E., et al.\ 1988, \apj, 326, 924 

\bibitem[{{Bogey et al.} (1987)}]{1987JMoSp.122..313B} Bogey, M., Demuynck, C., 
Destombes, J.~L., 
\& Dubus, H.\ 1987, Journal of Molecular Spectroscopy, 122, 313 




\bibitem[{{Caselli et al.}  (1999)}]{Caselli_1999} Caselli, P., Walmsley, C. M.,Tafalla, M., Dore, L., and Myers, P. C.,
1999, ApJ, 523, 165


\bibitem[{{Caselli et al.} (2002)}]{Caselli_2002} Caselli, P., Walmsley, 
C.~M., Zucconi, A., et al.\ 2002, \apj, 565, 344 

\bibitem[{{Cazaux et al.} (2011)}]{Cazaux_2011} Cazaux, S., Caselli, P., 
\& Spaans, M.\ 2011, \apjl, 741, L34 

\bibitem[{{Ceccarelli et al.} (1998)}]{Ceccarelli_1998} Ceccarelli, C., Castets, A., Loinard, L., Caux, E., \& Tielens, A.~G.~G.~M.\ 1998, \aap, 338, L43 

\bibitem[{{Ceccarelli et al.} (2007)}]{Ceccarelli_2007} Ceccarelli, C., 
Caselli, P., Herbst, E., Tielens, A.~G.~G.~M., 
\& Caux, E.\ 2007, Protostars and Planets V, 47 

\bibitem[{{Chandra 
\& Kegel} (2000)}]{Chandra_2000} Chandra, S., \& Kegel, W.~H.\ 2000, \aaps, 142, 113 

\bibitem[{{Cox et al.} (1987)}]{Cox_1987} Cox, P., Guesten, R., \& Henkel, C. 1987, A\&A, 181, L19

\bibitem[{{Crapsi et al.} (2005)}]{Crapsi_2005} Crapsi, A., Caselli, P., Walmsley, C. M., Myers, P. C.,
Tafalla, M., Lee, C. W., \& Bourke, T. L. 2005, ApJ, 619, 379


\bibitem[{{Dalgarno 
\& Lepp} (1984)}]{Dalgarno_1984} Dalgarno, A., \& Lepp, S.\ 1984, \apjl, 287, L47 

\bibitem[{{Gerin et al.} (1987)}]{Gerin_1987} Gerin, M., Wootten, H.~A., Combes, F.,
et al.\ 1987, \aap, 173, L1 

\bibitem[{{Guelin et al.} (1977)}]{Guelin_1977} Guelin, M., Langer, 
W.~D., Snell, R.~L., \& Wootten, H.~A.\ 1977, \apjl, 217, L165 




\bibitem [{{Herbst \& Leung} (1989)}]{Herbst_1989} Herbst, E., Leung, C. M., 1989, ApJ Supp., 69, 271

\bibitem[{{Huang 
\& Lee} (2011)}]{Huang_2011} Huang, X., \& Lee, T.~J.\ 2011, \apj, 736, 33 


\bibitem[{{Keto 
\& Caselli} (2010)}]{Keto_2010} Keto, E., \& Caselli, P.\ 2010, \mnras, 402, 1625 

\bibitem[{{Linsky et al.} (1993)}]{Linsky_93} Linsky, J.~L., Brown, 
A., Gayley, K., Diplas, A., Savage B. D., 1993, \apj, 402, 694 

\bibitem[{Lis et al. (2002)}]{Lis_2002} Lis, D. C., Roueff, E., Gerin, M., Phillips, T. G.,
Coudert, L. H., van der Tak, F. F. S., \& Schilke, P., 2002, ApJ, 571, L55

\bibitem[{{Lucas et al.} (2000)}]{Lucas_2000} Lucas, R., \& Liszt, H. 2000, A\&A, 358, 1069

\bibitem[{{Madden et al.} (1989)}]{Madden_1989} Madden, S. C., Irvine, W. M., Matthews, H. E., Friberg, P., \& Swade, D. A. 1989, AJ, 97,1403





\bibitem[{{Park et 
al.} (2006)}]{Park_2006} Park, I.~H., Wakelam, V., \& Herbst,
E.\ 2006, \aap, 449, 631 

\bibitem[{{Parise et 
al.} (2004)}]{Parise_2004} Parise, B., Castets, A., Herbst, E., et al.\ 2004, \aap, 416, 159 

\bibitem[{{Parise et al.} (2006)}]{Parise_2006} Parise, B., Ceccarelli, C., Tielens, A. G. G. M.,
Castets, A., Caux, E., Lefloch, B., \& Maret, S., 2006, A\&A, 453, 949

\bibitem[{{Parise et al.} (2009)}]{Parise_2009} Parise, B., Leurini, S., Schilke, P., Roueff, E.,
Thorwirth, S., \& Lis, D. C., 2009, A\&A 508, 737

\bibitem [{{Pety} (2005)}]{Pety_2005}Pety, J. 2005, in SF2A-2005, ed. F. Casoli, T. Contini, J. Hameury, \& L. Pagani,
EDP Sciences Conf. Ser., 721

\bibitem[{{Roberts et al.} (2000)}]{Roberts_2000} Roberts, H., \& Millar, T. J. 2000, A\&A, 364, 780

\bibitem[{{Roberts et al.} (2002)}]{Roberts_2002} Roberts, H., Herbst, 
E., \& Millar, T.~J.\ 2002, \mnras, 336, 283 

\bibitem[{{Roberts et al.} (2003)}]{Roberts_2003} Roberts, H., Herbst, E., \& Millar, T.~J.\ 2003, \apjl, 591, L41 

\bibitem[{{Roberts et al.} (2004)}]{Roberts_2004} Roberts, H., Herbst, E., \& Millar, T. J. 2004, A\&A, 424, 905

\bibitem[{{Roueff et al.} (2005)}]{Roueff_2005} Roueff, E., Lis, D.~C., van der Tak, F.~F.~S., Gerin, M., \& Goldsmith, P.~F.\ 2005, \aap, 438, 585 

\bibitem[{{Roueff et al.} (2007)}]{Roueff_2007} Roueff, E., Herbst, E., Lis, D. C., \& Phillips, T. G. 2007, ApJ, 661, L159

\bibitem[{{Savi{\'c} et al.} (2005)}]{Savic_2005} Savi{\'c}, I., 
Schlemmer, S., \& Gerlich, {\it}.\ 2005, \apj, 621, 1163 

\bibitem[{{Schnee et al.} (2007)}]{Schnee_2007} Schnee, S., Caselli, P., 
Goodman, A., et al.\ 2007, \apj, 671, 1839 

\bibitem[{{Sch{\"o}ier et 
al.} (2005)}]{Schoier_2005} Sch{\"o}ier, F.~L., van der Tak, F.~F.~S., van Dishoeck, E.~F., \& Black, J.~H.\ 2005, A\&A, 432, 369 


\bibitem[{{Shah 
\& Wootten} (2001)}]{Shah_2001} Shah, R.~Y., \& Wootten, A.\ 2001, \apj, 554, 933 

\bibitem[{{Spezzano et al.} (2012)}]{Spezzano_2012} Spezzano, S., Tamassia, F., Thorwirth, S., Thaddeus, P.,
Gottlieb, C. A., and McCarthy, M., 2012, ApJ Supp., 200, 1

\bibitem[{{Tafalla et al.} (1998)}]{Tafalla_1998} Tafalla, M., Mardones, 
D., Myers, P.~C., et al.\ 1998, \apj, 504, 900 

\bibitem[{{Taquet et al.} (2012)}]{Taquet_2012} Taquet, V., Ceccarelli, 
C., \& Kahane, C.\ 2012, \apjl, 748, L3 



\bibitem[{{Tielens} (1983)}]{Tielens_1983} Tielens, A.~G.~G.~M.\ 1983, \aap, 119, 177 

\bibitem[{{Tin{\'e} et 
al.} (2000)}]{Tine_2000} Tin{\'e}, S., Roueff, E., Falgarone, E., Gerin, M., \& Pineau des For{\^e}ts, G.\ 2000, \aap, 356, 1039 

\bibitem[{{Thaddeus et al.} (1981)}]{Thaddeus_1981} Thaddeus, P., Guelin, 
M., \& Linke, R.~A.\ 1981, \apjl, 246, L41 

\bibitem[{{Thaddeus et al.} (1985)}]{Thaddeus_1985} Thaddeus, P., Vrtilek, J. M., \& Gottlieb, C. A. 1985,
ApJ, 299, L63

\bibitem[{{Turner} (1990)}]{Turner_1990} Turner, B.~E.\ 1990, \apjl, 
362, L29 


\bibitem[{{van der Tak et al.} (2007)}]{Vandertak_2007} van der Tak, F.~F.~S., Black, J.~H., Sch{\"o}ier, F.~L., Jansen, D.~J., \& van Dishoeck, E.~F.\ 2007, A\&A, 468, 627 

\bibitem[{{Vrtilek et al.} (1987)}]{Vrtilek_1987} Vrtilek, J. M., Gottlieb, C. A. \& Thaddeus, P. 1987, ApJ,
  314, 716

\bibitem[Ward-Thompson et al. (1999)]{WardThompson_1999} Ward-Thompson, 
D., Motte, F., \& Andre, P.\ 1999, \mnras, 305, 143 

\bibitem[{{Wilson 
\& Rood} (1994)}]{Wilson_1994} Wilson, T.~L., \& Rood, R.\ 1994, \araa, 32, 191 


\end{thebibliography}
\end{document}